\documentclass[acmsmall]{acmart}

\AtBeginDocument{%
  \providecommand\BibTeX{{%
    \normalfont B\kern-0.5em{\scshape i\kern-0.25em b}\kern-0.8em\TeX}}}

\setcopyright{acmlicensed}
\copyrightyear{2024}
\acmYear{2024}
\acmDOI{XXXXXXX.XXXXXXX}

\acmConference[Conference acronym 'XX]{Make sure to enter the correct
  conference title from your rights confirmation email}{June 03--05,
  2024}{Woodstock, NY}
\acmISBN{978-1-4503-XXXX-X/18/06}

\usepackage{booktabs}
\usepackage{longtable}
\usepackage{float}
\usepackage{pdflscape}
\usepackage{multirow}
\usepackage{array}
\newcolumntype{P}[1]{>{\raggedright\arraybackslash}p{#1}}
\usepackage{graphicx}

\begin{document}


\title{Expanding Perspectives on Data Privacy:\\ Insights from Rural Togo}





\author{Zoe Kahn}
\affiliation{School of Information, University of California, \country{USA}
}
\email{zkahn@berkeley.edu}

\author{Meyebinesso Farida Carelle PERE}
\affiliation{
\institution{Independent Researcher}
\country{Togo}
}
\email{perecarelle@gmail.com}

\author{Emily Aiken}
\affiliation{
\institution{School of Information, University of California, Berkeley}
\country{USA}
}
\email{emilyaiken@berkeley.edu}

\author{Nitin Kohli}
\affiliation{
\institution{Center for Effective Global Action, University of California, Berkeley}
\country{USA}
}
 \email{nitin.kohli@berkeley.edu}

\author{Joshua E. Blumenstock}
\affiliation{
\institution{School of Information, University of California, Berkeley}
\country{USA}
}
 \email{jblumenstock@berkeley.edu}



\renewcommand{\shortauthors}{Anonymized for Review.}

\begin{abstract}
Passively collected ``big'' data sources are increasingly used to inform critical development policy decisions in low- and middle-income countries. While prior work highlights how such approaches may reveal sensitive information, enable surveillance, and centralize power, less is known about the corresponding privacy concerns, hopes, and fears of the people directly impacted by these policies --- people sometimes referred to as \textit{experiential experts}. To understand the perspectives of experiential experts, we conducted semi-structured interviews with people living in rural villages in Togo shortly after an entirely digital cash transfer program was launched that used machine learning and mobile phone metadata to determine program eligibility. This paper documents participants' privacy concerns surrounding the introduction of big data approaches in development policy. We find that the privacy concerns of our experiential experts differ from those raised by privacy and development \textit{domain experts}. To facilitate a more robust and constructive account of privacy, we discuss implications for policies and designs that take seriously the privacy concerns raised by both experiential experts and domain experts. 
\end{abstract}




\maketitle


\section{Introduction}
Over the past decade, governments in low- and middle-income countries (LMICs) have increasingly used passively collected ``big'' data to inform policy decisions. Machine learning algorithms now leverage satellite imagery \cite{jean2016combining}, mobile phone metadata \cite{blumenstock2015predicting}, and social media data \cite{fatehkia2020relative} in settings ranging from the targeting of humanitarian aid \cite{aiken2022machine,smythe2022geographic} to determining lending decisions \cite{bjorkegren2020behavior} and informing pandemic response \cite{peak2018population, oliver2020mobile}. This use of big data to inform critical development policy decisions has reinvigorated longstanding debates about data privacy: big data may improve decision making in certain settings, but they can also reveal sensitive information about people and communities \cite{de2013unique,taylor2015name,blumenstock2023big}. 

A wide range of \textit{domain experts} in sociology, development studies, computer science, privacy, and legal studies have raised empirical concerns around data privacy and re-identification \cite{de2013unique}, proposed data protection frameworks and best practices \cite{cobb2016computer}, analyzed power structures \cite{taylor2015name, taylor2016no, VanStaden2023LocalizsedTrust}, and studied data sharing practices \cite{Abebe_2021} in big data approaches to development. However, less research has engaged with the people directly impacted by these policies --- sometimes referred to as \textit{experiential experts} --- to understand their hopes for, and concerns about, the introduction of big data approaches to development. We use the term \textit{experiential experts} from Young et al. (2019) who define experiential experts as ``people who are living the experience or those closely associated with someone living the experience'' \cite{young2019toward}, a concept that has been expanded on by Magassa and Friedman (2024) in the context of the US criminal justice system \cite{magassa2024toward}. This provides a useful framework to help avoid false dichotomies often drawn between ``western'' and ``non-western'' societies or the ``global north'' and ``global south.'' These dichotomies often fail to capture important heterogeneity within each category, for instance by overlooking differences between urban and rural places, large and small communities, and affluent and poor people, among others.\footnote{When we shared a draft of this paper with one village leader, they noted that Africans --- and the African perspective --- have been influenced by Europeans (and to a lesser degree Americans) through the education system. This further challenges the false dichotomy often drawn between western and non-western values and perspectives.} Moreover, unlike the terms ``impacted people'' or ``impacted communities'' --- which suggest that people are passive receivers of technology --- the use of \textit{experiential experts} asserts that people have agency and legitimates their knowledge as a form of expertise. To that end, we conducted interviews with one group of experiential experts --- people living in rural villages in Togo --- to understand, from their perspective, what data privacy concerns do (and do not) arise when big data are used to inform development policy.

Togo is a country of roughly eight million in West Africa. During the COVID-19 pandemic, Togo launched \textit{Novissi}, one of the world's first entirely digital cash transfer programs, which provided mobile money transfers directly to people  who were experiencing extreme poverty in the 200 poorest cantons (admin-3 units roughly equivalent to counties in the United States) \cite{lawson2023novissi}. The \textit{Novissi} program was particularly novel because it used mobile phone metadata, processed via machine learning algorithms, to estimate poverty status and determine program eligibility \cite{aiken2022machine}. To receive a cash transfer, beneficiaries were required to have a sim card and voter ID registered to one of the 200 eligible cantons. The registration process took place using a USSD platform accessible on basic (2G) feature mobile phones. Eligible beneficiaries immediately received a cash transfer to their mobile money account (accessible on basic (2G) feature mobile phones); non-beneficiaries received a SMS informing them that they were not eligible at this time \cite{lawson2023novissi}. According to a 2018-2019 field survey, 50\% of individuals and 77\% of households in rural areas of Togo owned a mobile phone in 2018 \cite{aiken2022machine}. See Appendix \ref{app:mobilephonemetadata} for further discussion of mobile phone metadata and examples of how they are being used to inform public policy.

We conducted semi-structured interviews with people living in rural villages in Togo to understand their data privacy concerns associated with big data approaches to development policy, specifically related to uses of mobile phone metadata. We find that experiential experts and domain experts converge on their desire to enable ``helpful'' uses of big data and mitigate ``harmful'' uses. However, experiential experts and domain experts diverge in terms of the types of privacy harms they raise. In particular, while domain experts tend to emphasize concerns about surveillance or the centralization of power \cite{taylor2015name, taylor2016no, landau2016transactional, Abebe_2021}, our experiential experts principally raised relational privacy concerns tied to to shame, jealousy, and autonomy. 

This paper makes three main contributions. First, we document the privacy concerns of experiential experts in LMICs (specifically, people living in rural villages in Togo) surrounding the introduction of big data approaches to development policy. Second, we demonstrate that the privacy concerns most commonly raised by domain experts do not adequately account for the privacy concerns of our experiential experts. Third, working toward a more robust and constructive account of privacy in the case of big data approaches to development policy, we discuss implications for designs that take seriously the privacy concerns raised by both experiential experts and domain experts. 
\section{Literature Review}
Our work is situated within four main bodies of literature: privacy as sociotechnical; the relationship between privacy and poverty; common privacy concerns raised by domain experts about the use of mobile phone metadata to inform development policy; and methodological approaches to elicit data privacy concerns from experiential experts.
\subsection{Privacy as Sociotechnical (in Brief)}
The concept of privacy evolves in response to social, cultural, political, and technological transformations \cite{mulligan2016privacy, bajpai2017privacy, blumenstock2023big}, and is a dynamic boundary regulating process \cite{altman1975environment, altman1977privacy}, that depends on contextual norms \cite{nissenbaum2004privacy}. For example, the introduction of social media as a novel form of communication has raised new privacy concerns, especially for marginalized groups \cite{Abokhodair2016Privacy, Jack2019Cambodia}. Similarly, the proliferation of mobile phones in LMICs has been accompanied by a complex set of phone sharing practices \cite{Burrell2010PhoneSharing, Steenson2017Beyondpersonal, Ahmed2019PersonalStuff, Porter2020sharing, Aiken2022PhoneSharing, paudel2023deep} which has led people --- women in particular --- to adopt a range of techniques to maintain privacy and autonomy \cite{Burrell2010PhoneSharing, Ahmed2017PhoneSharing, Ahmed2019PersonalStuff, paudel2023deep}. Moreover, practices that are acceptable in one place or culture at a particular moment in time may raise privacy concerns in another \cite{Abokhodair2016Privacy, Srinivasan2018povertyprivacy}. To our knowledge, only one study has engaged experiential experts to understand the privacy harms that arise from the introduction of mobile phone metadata: in that case, university professionals in Wales were engaged as experiential experts around the use of mobile phone metadata in healthcare research \cite{jones2019public}. While many of the privacy harms raised in these examples are surfaced by experiential experts, privacy harms can be surfaced in a variety of ways. 

People with different expertise are often well-suited to surface different types of privacy harms. For example, people with technical expertise may be well-suited to identify security vulnerabilities \cite{de2013unique, landau2016transactional}; people with historical expertise may be well-suited to place new technologies within broader patterns of surveillance and control \cite{seltzer2001dark, parenti2003soft, browne2015dark}. Experiential experts may be well-suited to surface privacy harms from new technologies being integrated into their daily lives \cite{Gilliom2001Overseers, Srinivasan2018povertyprivacy, VanStaden2023LocalizsedTrust}. Consider the introduction of location tracking applications, for example: privacy and security experts (domain experts) raised concerns about corporate and government surveillance; meanwhile, women (experiential experts) raised concerns related to intimate partner violence \cite{Slupska2019IPV, Levy2020IPV}. Each expert brings a set of privacy concerns ``into view'' that may otherwise be ``out of view.'' 

\subsection{Privacy and Poverty}
When governments introduce new technologies to address poverty there are a set of privacy
considerations that are likely to arise. Here we discuss two: (1) the tension between legibility and surveillance and (2) shame, stigma, and jealousy. In general, in order to be eligible for anti-poverty benefits, an individual's identity and poverty status must be legible to the state. In practice, this often requires that beneficiaries provide detailed data on their lives to prove that they are ``deserving'' of benefits \cite{Scott1998SeeingLikeState, Gangadharan2017privacypoor, Eubanks2018AutomatingInequality, Srinivasan2018povertyprivacy}. While traditionally much of these data are self-reported, it is increasingly common for development actors to use passively collected administrative and digital data to measure poverty \cite{aiken2022machine}. People living amidst bureaucratic surveillance systems describe their experience in a variety of ways: some explicitly use the language of surveillance and a right to privacy \cite{ahmed2017privacy, Srinivasan2018povertyprivacy}, while others describe the day-to-day work that is required to navigate complex bureaucratic systems \cite{Gilliom2001Overseers} or the trusted relationships they develop with organizations (including the state) that develop over time \cite{VanStaden2023LocalizsedTrust}.

Beneficiaries of anti-poverty programs also raise a particular set of privacy concerns related to the shame and stigma of poverty \cite{roelen2020Shame}. Shame and stigma have been shown to limit the success of anti-poverty programs in both low- and high-income countries: program uptake often hinges on designing programs such that people are able to receive benefits while maintaining dignity \cite{Ewoudou2009Stigma, melvin2022paper, della2022selective}. Beyond shame and stigma, a fear of jealousy and envy can also lead people to conceal data that could indicate things going well; this may be particularly true among people living in poverty where small advantages can generate substantial envy \cite{clanton2006jealousy, della2022selective}.

\subsection{Common Data Privacy Concerns Raised by Domain Experts: Mobile Phone Metadata to Inform Development Policy}
\label{sec:domainexperts}
While mobile phone metadata has been useful in informing development policy, domain experts from 
sociology, development studies, computer science, privacy, and legal studies have studied privacy concerns that arise from the analysis of mobile phone metadata by governments and other development actors \cite{de2013unique, taylor2015name, taylor2016no, landau2016transactional, kohli2023privacy}. These concerns tend to focus on data privacy harms associated with the collection, aggregation, and use of mobile phone metadata by large, powerful actors such as governments, corporations, or researchers. Here we discuss four common privacy concerns raised by domain experts. See Appendix \ref{app:mobilephonemetadata} for a full description of mobile phone metadata and its use in development policy. 

\textit{Re-identification, reconstruction, and inference.} Mobile phone metadata is vulnerable to reconstruction, re-identification, and inference even after common data privacy strategies have been implemented (e.g., pseudonymization, using aggregate statistics, and not tying phone numbers to names or identities). Sophisticated attacks can re-identify mobile phone records \cite{sharad2013anonymizing} and reconstruct individual mobility traces \cite{chen2019complete}. Research has shown that mobility traces derived from mobile phone metadata are highly unique, such that aggregating statistics may still ``leak'' information about individual users \cite{de2013unique, pyrgelis2017does, xu2017trajectory}. The locations contained in these traces are also highly revelatory \cite{mayer2016evaluating, kohli2023privacy}, enabling inferences of political affiliations \cite{thompson2019twelve}, sexual preferences \cite{ovide2021}, and an individual's overall way of life \cite{2012us}.

\textit{Consent.} Mobile phone metadata and other passively collected data traces raise questions of consent \cite{solove2023murky}. Many research and policy projects involving mobile phone metadata have conducted analysis without the direct consent of data subjects \cite[e.g.][]{eagle2010network, frias2012relationship, hernandez2017estimating, aiken2022machi`ne}, as the data is often shared and controlled by mobile network operators and consent is often not required by local laws and/or research ethics review boards \cite{taylor2016no}. Taylor (2016) identifies that connectivity, literacy, and dynamic updating can make consent particularly challenging in the context of mobile phone metadata \cite{taylor2016no}.

\textit{Compliance with privacy law.} Mobile phone metadata and its use to inform development policy may challenge existing legal frameworks. Many LMICs do not have specific data privacy regulation governing the use of mobile phone metadata. US and EU privacy law have typically drawn a distinction between the analysis of communication content versus communication metadata \cite{conley2014metadata, landau2020categorizing}; however, some scholars argue that mobile phone metadata is so revelatory that it does not fit squarely within the definition of communication metadata, particularly given concerns about individual and group privacy \cite{letouze2015law, landau2016transactional}. In international law, privacy is a derogable right\footnote{See Articles 4 and 17 of the International Covenant on Civil and Political Rights.}: certain aspects of privacy (e.g., informed consent) can be infringed upon in certain situations (e.g., crisis, conflict) so long as the use is necessary for a legitimate legal purpose and proportionate for the task at hand \cite{rights1996international, effnecessary, kohli2021leveraging}.

 \textit{Surveillance.} Mobile phone metadata could be used to centralize authority in a surveillance state. Taylor (2016) identifies that even well-intentioned uses of mobile phone data for development risk exposure to ``function creep'' in which uses of mobile phone data expand beyond initial objectives, and potentially lead to nefarious and repressive uses such as targeting political dissidents or repressing migration \cite{taylor2016no}.

\subsection{Methods to Elicit Data Privacy Concerns From Experiential Experts}
There are a variety of established methods to elicit the privacy concerns of experiential experts. In this research, we use common interview techniques and extend scenario based approaches. We use scenario based methods to overcome several of limitations of related methodological approaches. Specifically, survey based methods to elicit privacy preferences --- while well-suited to evaluate privacy perceptions at scale that allows for geographic comparisons --- are often hindered by lack of context and are vulnerable to the privacy paradox \cite{barth2017privacy} (and the reverse privacy paradox \cite{colnago2023there}). Further, interviews with people about how they manage privacy in everyday interactions with ``big'' data systems --- while highly contextual --- require that people be aware of the data being collected and are limited by the here and now \cite{Abokhodair2016Privacy, Srinivasan2018povertyprivacy, Jack2019Cambodia}. Scenario based approaches (which leverage short narratives that place technological systems alongside humans and their activities \cite{carroll2003making}), on the other hand, allow participants to reflect on privacy in context without being bound to prior awareness of the data collected or the here and now. Scenarios have been used to elicit privacy concerns in a variety of contexts including: learning analytics in universities \cite{li2021s, li2022scenario}, public health surveillance \cite{seberger2021post}, online e-commerce \cite{ackerman1999privacy}, data disclosure using apps \cite{seberger2021empowering}, government use of open data platforms \cite{ruijer2017connecting}, privacy expectations on mobile applications \cite{martin2016putting}, women’s health \cite{bardzell2019re}, and automated gender recognition \cite{hamidi2018gender}. Across these various research contexts, scenarios have primarily been used to elicit privacy concerns in response to open-ended scenarios, particularly from participants located in the United States and Europe. 
\section{Methods}
\label{sec:methods}

\subsection{Methods Overview}

During the period between January 2021 and April 2023, we conducted participant observation and semi-structured interviews with a total of 124 individuals in rural villages in Togo, as part of a broader research agenda focused on the use of mobile phones and mobile phone metadata in development policy. Of those 124 subjects, 32 were asked specifically about data privacy. Those 32 subjects are the primary focus of this paper, and their responses form the basis for the findings we present. However, at various points in the manuscript, and particularly when describing our methods, we refer to the full sample of 124 subjects, as the full sample contributed to our understanding of local norms and values in rural Togo.

Our interactions with research subjects were divided into three phases: two pilot phases (\textit{Phase 1} and \textit{Phase 2}), and the primary data collection phase (\textit{Phase 3}) — see Table \ref{table1} for details. We describe these phases in turn below. All interviews took place in local languages (Ewe, Kabyé, Kotokoli, or French) with a Togolese interpreter.  

\subsection{Phase 1: Pilot Phone Interviews (January - May 2021)}

To begin to understand the research space, in \textit{Phase 1} we conducted phone interviews with 16 people living in rural villages across Togo. Participants were recruited from several larger national-scale phone surveys that members of our team conducted as part of a different project in 2020-2021. Collectively those surveys had 28,902 respondents, of whom 28,174  indicated that they would be interested to be contacted in the future about related research. From those 28,174, we sampled 16 participants stratified by gender, age, and whether or not they received a cash transfer from \textit{Novissi}; we also sought a representative sample of regions (north vs. south) and language (which tends to correlate with ethnicity). 

\textit{Method description: Phase 1.} The phone interviews lasted between 25 minutes - 3 hours each, and had two parts. In the first part, participants were asked about their experience with \textit{Novissi}. In the second part, participants were asked about the sensitivity of mobile phone metadata. Specifically, we asked how participants would feel if different pieces of mobile phone metadata (i.e., location of phone calls and text messages, number of phone calls and text messages, mobile money account balance) were accessed and/or used by different entities (i.e., government of Togo, researchers in the United States, researchers in Togo, NGOs, privacy companies, and/or for the purpose of determining eligibility for a cash aid program). These data privacy questions were inspired by the privacy analytic developed by Mulligan et al. (2016) \cite{mulligan2016privacy}.

\textit{Method learnings: Phase 1.} In these phone interviews, we observed that many participants were able to speak at length about their experience with \textit{Novissi}, but struggled to provide meaningful input on the questions related to data privacy. In large part, this arose because participants did not have a clear understanding of the data that is (and is not) recorded in mobile phone metadata. For example, some participants were confused by the granularity of location data and others (incorrectly) assumed that the content of phone calls and/or text messages was also recorded. This indicated that we needed to adjust our methods to provide participants with an understanding of the data that is (and is not) recorded in mobile phone metadata. 

\begin{table}
\caption{Overview of data collection location, mode, and goal by research phase \label{table1}}
\begin{tabular}{@{}P{2.8cm}P{7.8cm}P{2.2cm}@{}}
\toprule
\textbf{Phase}                                 & Data collection location, mode, and goal                                                                                                                          & \textbf{Number of participants} \\ \midrule
\textbf{Pilot Phase 1: Phone interviews}       & Phone interviews with people in rural villages across Togo to understand broad research space                                                                      & \textbf{16 total}               \\ \\
\textbf{Pilot Phase 2: Fieldwork}              & In--person fieldwork in two rural villages in Togo (one in south, one in north) to refine methods, pilot fieldwork logistics, and understand broad research space & \textbf{20 total}               \\ \\
\textbf{Phase 3: Fieldwork}   & In-person fieldwork in three rural villages (two in north, one in south) to conduct participant observation and interviews                                        & \textbf{88 total}               \\ 
                                               & \hspace{1cm} \textit{Topic: Data privacy}                                                                                                                            & \textit{(32)}                   \\
                                               & \hspace{1cm}  \textit{Topic: Novissi and digital social protection}                                                                                                   & \textit{(56)}                   \\ \\
\textbf{Total interviews (Phases 1, 2, 3)} &                                                                                                                                                                   & \textbf{124 combined total}     \\ \bottomrule
\end{tabular}
\end{table}

\textit{Data analysis: Phase 1.} Throughout \textit{Phase 1}, we iterated between data collection and memo writing. Although the methods were still being refined, memos from these initial phone interviews surfaced that participants were not especially concerned about the data privacy concerns commonly raised by domain experts related to surveillance and the misuse of data by powerful actors who are far away. 

\textit{Preparing for Phase 2: method refinement.} In reflecting on the phone interviews conducted in \textit{Phase 1}, we realized that in order to enable people living in rural villages in Togo to provide meaningful input on the privacy discussion taking place among domain experts, we needed a way to provide participants with varying levels of formal education and literacy with an understanding of the data that is (and is not) recorded in mobile phone metadata. To do so, in between \textit{Phase 1} and \textit{Phase 2}, we developed culturally relevant visuals to help scaffold conversations about mobile phone metadata (see Figure \ref{fig:all_images} for a description and image of the visuals in practice). 

 \subsection{Phase 2: Pilot Fieldwork (December 2022)}
To refine the methods and better understand local context, in \textit{Phase 2} we conducted 2 weeks of pilot fieldwork in two rural villages in Togo (one in the north, the other in the south). In this phase, we conducted in-person interviews with 20 people. 

\textit{Methods description Phase 2. }To pilot the visuals in context and refine interview questions, during \textit{Phase 2} we used the visuals in interviews to scaffold conversations about the sensitivity of different types of data that are (and are not) recorded in mobile phone metadata. In terms of interview questions, inspired by Future Workshops \cite{kensing2020generating}, we asked participants to tell us how they would like these data to be used, as well as how they would not want these data to be used. 

\textit{Method learnings: Phase 2.} We found that the visuals were effective at providing participants with a clearer understanding of mobile phone metadata. However, the interview questions that followed were difficult for many of our participants to answer. Many participants told us that they could not imagine how mobile phone metadata could be used for either good or bad purposes. This indicated that we needed to adjust the questions we were asking. 

\textit{Data analysis: Phase 2. } We continued to iterate between data collection and memo writing. Although the methods were still being refined, the pilot fieldwork continued to suggest that participants were not especially concerned about the data privacy concerns raised by domain experts related to surveillance and the misuse of data by powerful actors who are far away. Instead, we started to learn that participants may instead be concerned if people nearby (e.g., spouse, household, villagers) accessed and/or used mobile phone metadata. 

\textit{Preparing for Phase 3: method refinement.} In reflecting on the fieldwork and interviews conducted in \textit{Phase 2}, we realized that we were asking experiential experts to weigh in on a set of questions that did not leverage their unique expertise and instead were better suited to privacy or development domain experts. In the same way that development scholars may be uniquely positioned to comment on the (mis)use of data across history (and less on specific technical vulnerabilities), and privacy scholars may be uniquely positioned to comment on the specific technical vulnerabilities that arise from mobile phone metadata (and less on the (mis)use of data across history), we needed to ask questions that would enable experiential experts to leverage their unique expertise. Upon returning to the United States, we updated the interview guides accordingly.

\subsection{Phase 3: Fieldwork (February - April 2023)}
\subsubsection{Data Collection Overview}
In \textit{Phase 3}, we conducted 7 weeks of fieldwork in three rural villages. As part of the broader research agenda we conducted a total of 88 in-person interviews. Of those 88 subjects, 32 were asked specifically about data privacy (see Table \ref{table2} for a summary of the demographic characteristics of these participants). 

In addition to the semi-structured interviews, we conducted participant observation in all three villages. This included attending funerals; going to local markets; farming tomatoes, peanuts, and yams; seeing the process of distilling local palm liquor; spending hot afternoons under mango trees; and so forth. Data collection also included extended conversations with interpreters in the field who helped us understand local context and navigate the situated on the ground experience. Data was iteratively analyzed using an inductive qualitative approach.   

\begin{table}[]
\caption{Summary of participant demographics for participants interviewed about data privacy \label{table2}}
\begin{tabular}{ll}
\toprule
\textbf{Demographic}                                                                       & \textbf{Participants} \\ \midrule
\multicolumn{2}{l}{\textit{Panel A: Novissi beneficiary status}}                                                   \\
Beneficiary                                                                                & 15                    \\
Registered but not receive                                                                 & 13                    \\
Did not register                                                                           & 4                     \\ 
\textbf{Total}                                                                             & \textbf{32}           \\ \\
\multicolumn{2}{l}{\textit{Panel B: Occupation}}                                                                   \\
Farmer                                                                                     & 19                    \\
Sellers at local markets                                                                   & 5                     \\
Other (e.g. welder, tailor, retired, unemployed) & 8                     \\ 
\textbf{Total}                                                                             & \textbf{32}           \\ \\
\multicolumn{2}{l}{\textit{Panel C: Age}}                                                                          \\
35 or younger                                                                              & 13                    \\
Older than 35                                                                              & 18                    \\
Not reported                                                                               & 1                     \\ 
\textbf{Total}                                                                             & \textbf{32}           \\ \\
\multicolumn{2}{l}{\textit{Panel D: Education}}                                                                    \\
None                                                                                       & 9                     \\
Completed part or all of primary school                                                    & 7                     \\
Completed part or all of middle school                                                     & 16                    \\
Completed part or all of high school                                                       & 0                     \\
Completed part or all of university                                                        & 0                     \\ 
\textbf{Total}                                                                             & \textbf{32}           \\ \bottomrule
\end{tabular}
\end{table}

\subsubsection{Village Selection, Recruitment, and Access}
We recruited cantons (admin-3 units roughly equivalent to countries in the United States) and then coordinated with local leadership to select and recruit villages (See Appendix \ref{app:villagedescription} for a description of the three villages). Cantons were purposefully sampled for range and diversity along five characteristics: region (north/south), language (Ewe, Kabyé, Kotokoli), ethnicity (Ewe, Kabyé, Kotokoli, Peule/Fulani), \textit{Novissi} targeting method (individual-level targeting vs. regional-level targeting), and percent of eligible voters who registered for \textit{Novissi} (high/low). The second author --- who is Togolese --- helped ensure villages were recruited in a way that respected local norms. 

The research was introduced to local communities as a study to inform the design of future cash transfer programs (in Togo and beyond) that use mobile phones and digital technology. We emphasized that participation in the research would have no impact on whether or not any village or individual received aid today or in the future. Despite our best efforts, participants often thought --- at least initially --- that we were associated with \textit{Novissi}. We imagine that this perception is one reason participants talked about finances (particularly living poverty) with such prevalence, and may be one reason that participants often perceived the government to be an entity that would use data to provide financial support to people living in poverty. While we don't believe this perception would change the overarching conclusions of this paper, it may influence some of the specific examples provided.

\subsubsection{Participants}
 In all three villages, recruitment was facilitated through the village or canton Comité Villageois de Développement (CVD): a local role responsible for administrative work. We asked to interview people who would provide a range of perspectives, in particular seeking diversity in terms of: gender, age (35+ and 35-), and \textit{Novissi} beneficiary status (beneficiary, registered but not eligible to receive a cash transfer, did not register). Participants had varying levels of formal education, literacy, numeracy, and digital literacy. Participants were not recruited based on whether or not they owned a mobile phone (basic 2G feature phone or smartphone), though participants varied in terms of phone ownership. Some participants owned their own basic (2G) feature phone, others owned their own smartphones, and others did not own their own phone. Because we have ethnic and geographic diversity in our sample we believe our participants are varied in terms of political affiliation and support, though we did not directly ask about political affiliation.\footnote{Ethnicity and geography tend to correlate with support for the government in Togo \cite{OECD2015voting}.} See Table \ref{table2} for a summary of the 32 participants interviewed on the topic of data privacy and  Appendix \ref{app:participant_table} for a summary of the demographics for all 124 participants.\footnote{We do not include a participant table with demographic information on each participant to protect participants' identity.} 

\subsubsection{Interview Guide Description}
\label{interviewguide}

The interviews reported on in this article specifically asked about data privacy and did not mention \textit{Novissi} (specifically) or digital cash transfer programs (generally).\footnote{Exceptions include: (1) Interviews with two participants began by asking about \textit{Novissi} but the participants shifted the interview to discussing data privacy; (2) several participants brought up \textit{Novissi} themselves, in the context of the interviews on data privacy; and (3) several participants asked questions about \textit{Novissi} toward the end of interviews.} Interviews consisted of three sequential parts. Of the 32 participants, all (100\%) were asked questions for Part 1. Of these 32, 15 (46\%) also completed Part 2. And of these 15 participants, 13 (86\%) also completed Part 3 (See Table \ref{table3}).  

There are two reasons why all participants did not receive all three parts of the data privacy interview guide: one practical, the other methodological. Practically, many of the participants stopped their regular work activities to be interviewed. To respect their time, in some cases we only had time to discuss one or two of the interview parts. Methodologically, qualitative research aims to reach saturation: a point when no new information is learned. Different scopes of research (and interview questions) may take more or less data to reach saturation: a narrowly scoped question where many different people tend to feel similarly may reach saturation more quickly than a more broadly scoped question where there are a variety of different opinions \cite{lareau2021listening}. In our case, Part 1 of the data privacy interview guide was the most open-ended and elicited the widest variation in participant responses (which meant that it took longer to reach saturation). Part 2 of the data privacy interview guide was somewhere in the middle with respect to question scope; and Part 3 was the most narrowly scoped. For Part 2 and especially Part 3,  participants tended to respond quite similarly, which meant we reached saturation more quickly. We provide a brief description of each part of the data privacy interview guide below.

\begin{table}[]
\caption{Overview of Phase 3 data privacy interview by part and participants. As detailed  in Section \ref{interviewguide}, Part 1 addressed local norms and expectation of privacy, Part 2 sensitivity of mobile phone metadata, and Part 3 asked about four specific data privacy concerns raised by domain experts about the use of mobile phone metadata. \label{table3}}
\begin{tabular}{@{}ll@{}}
\toprule
\textbf{Data privacy interviews by part} & \textbf{Number of participants (N=32)} \\ \midrule
Part 1                               & 32 of 32                             \\
Part 2                               & 15 of 32                           \\
Part 3                               & 13 of 32                                                \\ \bottomrule
\end{tabular}
\end{table}

\textit{Part 1: local norms and expectation of privacy} asked open-ended questions to understand the type of data people felt was sensitive, why, and from whom. Interview questions were informed by the privacy analytic developed by Mulligan et al. (2016) \cite{mulligan2016privacy}.

\begin{figure}
\Description[]{}
\caption{Left: Photograph of visuals used to facilitate conversations with participants about four pieces of data recorded in mobile phone data: mobile money and mobile money transactions, phone calls, text messages, and location. Each visual is printed on card stock and laminated to facilitate easy rearrangement and reuse. The second author who is Togolese helped ensure the visuals were culturally relevant. For example, the image of the map is the map that is commonly taught in schools in Togo; the images of basic feature phones were used over smartphones because these more closely resemble the phones used by our participants in rural villages in Togo. Right: Photograph of visuals being used during an interview. Participants would often pick up, arrange, or point to the visuals during interviews to help articulate their ideas.}
\centering
\includegraphics[width=0.45\textwidth]{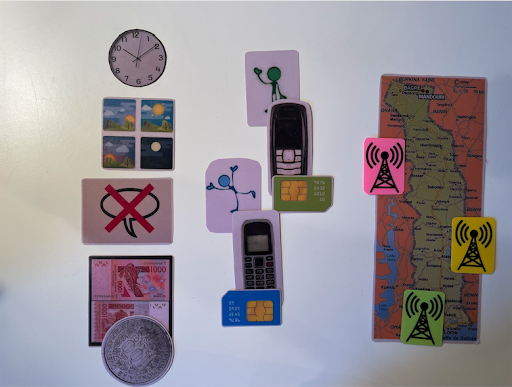}
\includegraphics[width=0.416\textwidth]{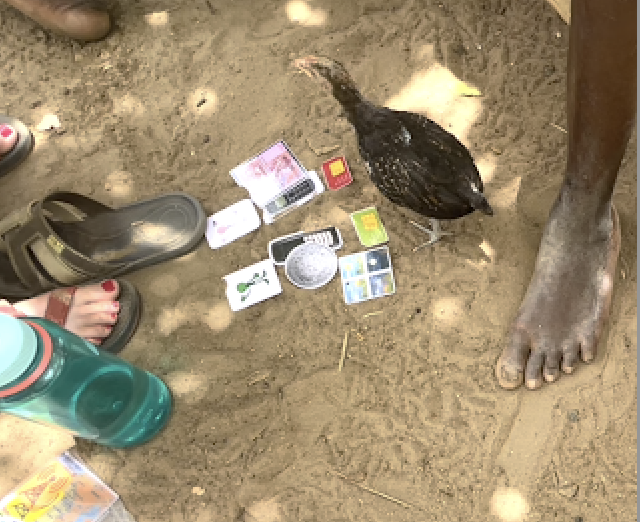}
\label{fig:all_images}
\end{figure}

\textit{Part 2: sensitivity of mobile phone metadata} used culturally relevant visuals to scaffold conversations about mobile phone metadata, asking questions to understand the sensitivity of different pieces of data recorded in mobile phone metadata, why, and from whom (see Figure \ref{fig:all_images} for a description and image of the visuals). Pieces of data included: records of mobile money and mobile money transactions, phone calls, text messages, and location. Interview questions were informed by the privacy analytic developed by Mulligan et al. (2016) \cite{mulligan2016privacy}. 

\textit{Part 3: four data privacy concerns raised by domain experts about the use of mobile phone metadata} extended existing scenario based approaches to elicit privacy concerns (e.g., \cite{carroll2003making, li2021s, li2022cross}). We used scenarios to enable experiential experts to weigh in on four specific privacy concerns raised by domain experts to argue for --- or against --- the use of mobile phone metadata in development. If we had not asked explicitly about the privacy concerns raised by domain experts, we would not have known if experiential experts merely talked about a different set of privacy concerns, or if they actually had a different perspective entirely. We developed four scenarios (see Appendix \ref{app:scenarios} for the scenarios). While all scenarios were not real, they were all possible \cite{blumenstock2015predicting, blumenstock2018estimating, aiken2022machine, aiken2023program, callen2023measuring} and inspired by the literature \cite{torkelson2020collateral, seltzer2001dark}. To provide a balanced account, two scenarios were intended to demonstrate ``helpful'' uses and two were intended to demonstrate ``harmful'' uses. The scenarios were read aloud to participants during the interview. After reading each scenario out loud, we asked participants how they felt about the government using mobile phone metadata for this purpose and why. The scenarios did not indicate how likely (or not) the scenario was to occur, and we did not ask participants how likely (or not) they perceived each scenario to be. While their perception of likelihood may influence the relative conviction of their response, we do not believe that it would influence the overarching sentiment.

We extended scenario based approaches for eliciting privacy concerns with respect to \textit{purpose} and \textit{regional and demographic context}. In terms of purpose, many scenarios are written in ways that intentionally leave the outcome ambiguous (e.g., \cite{martin2016putting, seberger2021empowering, seberger2021post, li2021s, li2022cross}). In contrast, we used scenarios to explore specific privacy concerns raised by domain experts --- in practice, the scenarios we developed were explicit about the intention of the government, as well as the (harmful or helpful) outcome. In terms of regional and demographic context, we broadened the geographic contexts of scenario based approaches to elicit privacy from people located primarily in urban areas in the United States or Europe \cite{martin2016putting, seberger2021post, li2021s, li2022cross} to people in rural West Africa with varying levels of formal education and literacy. Oral storytelling has a long history in Africa generally \cite{Abebe_2021}, and the second author who is Togolese informed us that oral storytelling remains a common practice in Togo, particularly in rural villages; we observed that reading the scenarios aloud to participants was both a format that people were familiar with and gravitated towards.

\subsubsection{Interpretation and Translation}
The following steps were taken to ensure that interpretations and interview transcript translations were accurate. Prior to conducting interviews, we spent 2-5 hours with each interpreter to review each interview guide and discuss the goals of the research. In several cases, we discussed important linguistic nuances. For example, it was important to communicate that data was ``recorded'' by mobile network operators as opposed to being ``known,'' which could suggest individual humans at the mobile network operator were reviewing the recorded data. More broadly, interpreters were encouraged to interpret for similarity of meaning (as opposed to similarity of words) because words may have different meanings in different languages and cultures \cite{van2010language}. To verify the quality of the transcription translations, all interviews were transcribed and translated into English. The second author who is Togolese and speaks French and Ewe (two of the four language interviews were conducted in) spot-checked transcriptions and translations. For interviews in Kotokoli and Kabyé, a local Kotokoli and Kabyé speaker spot checked the transcriptions and translations respectively. In several cases where an accurate translation was particularly important to understand meaning and/or nuance, we left comments in the transcripts to confirm the translation of participant quotes with a local speaker.

\subsubsection{Research Ethics}
This research was approved by the university Institutional Review Board (IRB). Due to varying levels of literacy among our participants, verbal informed consent was obtained prior to conducting interviews. Interviews took place at a location selected by each participant. Many participants chose a public place.\footnote{The second author, who is Togolese, told us that participants may prefer to have interviews in public because meeting in private may appear secretive and make participants vulnerable.} To ensure that participants felt that their participation was voluntary, given that interviews often took place in a public location and recruitment was often facilitated through a local leader, all participants were given the option to have an off the record conversation so that it appeared publicly that they were participating in the research even though their data would not be included in the research. No participant took us up on this offer. Furthermore, because local interpreters were often better suited to pick up on local social cues than the US-based research team, we often relied on interpreters to tell us if/when participants seemed uncomfortable and/or did not want to participate. In these cases, we thanked the participant for their participation and stopped the interview. For example, one participant was visibly ill and had recently been to the hospital so we stopped the interview after 15 minutes and thanked the participant for his time.

\subsubsection{Data Analysis}
Interviews were inductively coded iterating between open coding, focused coding, and memo writing. The second author, who is Togolese, provided key guidance to ensure that the findings and analysis are rooted in local interpretations. A draft of this paper was shared with the CVD from each of the three villages. At their request, we translated the article into French so they could review the paper and had several subsequent phone calls with each CVD to discuss the findings, answer questions, and seek feedback.
\section{Findings}
\label{sec:findings}
We begin by exploring participants' general privacy norms and expectations, not necessarily specific to digital data. We find that many of the privacy concerns discussed by our participants were relational, often tied to shame, jealousy, and limited autonomy. We then investigate concerns specific to the availability and use of mobile phone metadata, and highlight how the relational privacy concerns extend to this new type of data. With the exception of findings related to autonomy, we did not observe differences based on participants' age, gender, ethnic group identity, religion, village, or region.  

For each theme, we report the number of participants who raised that theme. All themes emerged inductively, without prompting. In other words, we did not ask participants to explicitly comment on shame, jealousy, autonomy, and/or distance, but rather participants surfaced these themes on their own. In general, the absence of a theme simply indicates that the theme did not surface in the interview, rather than that the participant actively disagreed. 

\subsection{Privacy in Rural Villages in Togo}
\label{sec:general_privacy}
In the first part of the interview, we asked participants (N = 32) to talk freely about what data in their lives felt sensitive: what they would not want others to know. Many participants said data that could indicate that their behaviors deviated from established social norms would be sensitive. Financial data --- and inferences related to financial status --- were most commonly mentioned, but similar themes surfaced for data that could be used to infer infidelity, infertility, stealing, witchcraft, and sexual relations. The revelation of financial data could bring shame (N = 14 of 32), make others jealous (N = 9 of 32), or limit their autonomy (N = 6 of 32). Yet, many participants said that the same data --- if shared with far away entities like the government of Togo --- did not raise the same concerns (N = 7 of 32). We focus this section on the sharing of financial data given its prevalence in our data. All names are pseudonyms. \footnote{In Togo, names often correspond with gender and/or ethnicity. Pseudonyms were selected to reflect the gender and ethnic identity of the participant.}     

\subsubsection{Shame in Hard Times}
Living in the poorest rural villages in Togo often means living on the brink of financial insecurity. Many of the participants we interviewed talked about moments when they were not able to provide food for their family, pay medical bills, or otherwise meet basic needs. These moments of financial hardship were described as shameful: so much so that they would not want people in their household or village to know. For example, Abide (a 38 year old woman) recalled a time a few years ago when she was unable to feed her family for three days and told no one except her brother-in-law. When asked why she chose to tell no one, she said, ``I could not tell my family that there was no food. It’s shameful... if I had told my family, they could have told me to leave my husband.'' Senyo (a 33 year old man) said he, ``wouldn't be able to look people in the eye and may have to leave the area'' if people in his village found out that he was not able to purchase a motorcycle and instead had a work for pay arrangement. In these interviews and others, shame was explicitly mentioned, experienced psychologically, and could lead to material consequences such as divorce or banishment. These findings are consistent with prior work discussing how people often experience poverty as shameful, in both ``western'' and ``non-western'' countries \cite{Ewoudou2009Stigma, roelen2020Shame, melvin2022paper}. 

\subsubsection{Jealousy in Prosperous Times}
In contrast to shame --- which arose in times of hardship --- concerns about jealousy arose during moments of relative financial prosperity. Many of our participants said that they did not want others in their household or community to know when finances were ``good'' because others might become jealous and invoke witchcraft. For example, Sadjida (a woman who did not report her age) said that she was careful what she tells people because, ``people may hear about your success, that you are managing to eat at home, and someone might become jealous and want bad things for you. They’ll use witchcraft so that you don’t prosper.'' Participants also talked about being cautious sharing their future financial plans. For example, Mawuena (a 35 year old man) said that if he were planning to build a new house he would not tell his in-laws because they may talk to others who ``may not have enough to eat today. The envy will start and they will try to ruin [my plans to build a house] before it even starts. So, I will finish [buying the land and building the house] before telling my in-laws.'' As shown in the quotes above, jealousy was often explicitly mentioned by name in interviews, tied to witchcraft, and could limit what people shared when finances were comparatively good. These findings are consistent with prior research on the sociology of emotions that documents how, in some situations, people conceal pieces of data that could indicate financial successes to prevent people from becoming jealous, particularly in poorer communities where small differences can lead to inequality \cite{clanton2006jealousy}. Moreover, prior work similarly finds a connection between jealousy and witchcraft in Africa, as well as other LMICs \cite{leistner2014witchcraft, geschiere2013witchcraft}. 

\subsubsection{Limited Autonomy and Financial Control}
While shame and jealousy were emotions that limited what participants shared with those in their household or village, autonomy was referenced as the reason people chose not to share finances with their spouse. Several of the men and women we interviewed said that if ``the marriage was good'' a couple would share finances with one another; but if ``the marriage was bad'' they would not. For example, the day before we interviewed Tchilabalo (a 36 year old man) we arrived at a neighboring house with the intention of interviewing another man. When we arrived, there were perhaps 10 adults sitting in a circle on wooden benches under a large mango tree. As soon as we arrived, our interpreter informed us that it was not a good time to conduct the interview. The following day when we interviewed Tchilabalo he told us what had occurred: the man we had intended to interview had been saving money to pay off a microfinance loan. The previous night, the man and his wife got into a large disagreement where the man beat his wife. The following night, the wife stole the money and left in the middle of the night. Tchilabalo drew on this experience to explain why women should not know the finances of men. 

Women were similarly concerned that men would take their money or use it frivolously. For example, Meyebinebè (a 42 year old woman) said that she never shares her finances with her husband because he would ``go drinking and come back and beat me, slap me harshly...the man can ask for a loan but he will never pay back the money. Let’s suppose you are making a profit of 1,500 FCFA [the currency used in Togo] from your business and tell your husband. He will come and ask you to give him 1,000 FCFA. And suppose you spend the remaining 500 FCFA on cooking, now what will be your profit? You are surely losing. That’s why we women keep quiet about our financial information.'' During the interview, Meyebinebè briskly stood up as she described earning a profit to indicate that she was ``succeeding,'' and then proceeded to forcefully sit down when she described how her husband could limit her success. 

More generally, both men and women we interviewed were acutely aware that sharing financial data with a spouse could limit their autonomy to control finances, particularly if there was tension in the relationship. These findings broadly align with prior work documenting how men and women receive pressure to share money with friends and family (which economists often refer to as ``kinship tax''), however our participants most commonly talked about this pressure arising with their spouse, as opposed to broader social pressure \cite{squires2018kinship, riley2022resisting, jakiela2016does}. Moreover, these findings are aligned with literature documenting various strategies men and women use to conceal finances to maintain autonomy \cite{Burrell2010PhoneSharing, jakiela2016does, castilla2019s, jeanfreau2020our}. 

\subsubsection{Physical and Social Distance as a Kind of Privacy Protection}
The above results show how shame, jealousy, and autonomy regulate what people share about their finances with others nearby: spouses, people in the household, or people in the village. In interviews, when we asked participants if they would be concerned if the same financial data was shared with the government; participants generally said this would not be a concern. They often responded that the government was far away physically and --- perhaps more importantly --- socially: suggesting that, unlike people nearby, the government could not make someone feel shame, become jealous, or limit an individual's autonomy. Instead, participants often raised the hope that the government might be able to step in to provide assistance if provided with data on their financial status. This is a different framing of distance than is common among the privacy concerns raised by domain experts, which typically sees distance as a problem because it centralizes power and enables surveillance \cite{taylor2015name, Abebe_2021}. 

Many of the participants who were concerned about sharing financial data with people close by said that they would be fine --- even happy --- if the same data was shared with the government. After Abide told us that she did not tell anyone aside from her brother in-law about the three day period where her family went without food, she also stated, ``I won’t be ashamed if the government knows about it because they are not here so they won’t talk about you; but if they [are close to you and] know [you], they will talk about you...and the government can provide support.'' Holali (a 30 year old woman) who previously told us she would not want people in her village to know that she borrowed money from her father in-law, said that she would ``feel nothing'' if the government knew this because, ``I think the government is normally there to assist people financially, so if they know such information about me, it is nothing.'' Tchilalo (a 28 year of woman) similarly viewed the government as a positive actor that could provide assistance during difficult times (although she was adamant that her husband should not know her finances because he might ask for a loan and not pay it back). Tchilalo also felt that people in her household and village should not know about her finances because it might bring shame to her husband if they found out that she was paying for items that her husband should be paying for; however, she would feel ``good'' if the government had access to this data because, ``they can know who is the poorest so if there is an aid program, they can decide those who should be prioritized.'' 

Beyond the government, we often found that participants told us --- researchers who are physically and socially distant from the community --- data that could reveal information that they often withheld from people nearby. For example, Abide said that aside from her brother-in-law, we were the only people she had told about this difficult three-day period where her family went without food. This is consistent with experiences of other qualitative researchers who, as outsiders to the community, have limited access to some information and interactions, but gain access to other information that might not be shared within the community, as participants are less concerned about their public image \cite{lareau2021listening}. 

\subsection{Privacy and Mobile Phone Metadata}
In the second part of the interview, we asked participants (N = 15) about the sensitivity of mobile phone metadata. To scaffold these conversations, we used culturally relevant visuals that enabled participants to respond to four specific pieces of data that are, and are not, recorded in mobile phone metadata (see Appendix \ref{app:mobilephonemetadata} for a description of data recorded in mobile phone metadata; see Figure \ref{fig:all_images} for images of the visuals). Mirroring the general privacy concerns documented in Section \ref{sec:general_privacy}, we found that the privacy concerns arising from the introduction of mobile phone metadata centered around the possibility that the data --- or inferences drawn from the data --- could lead to relational harms such as shame (N = 4 of 15), jealousy (N = 6 of 15), or limited autonomy (N = 11 of 15). Likewise, participants were primarily concerned when the data --- or sensitive inferences drawn from that data --- was shared with people nearby (i.e., spouse, household, village), rather than the collection, aggregation, and use of the data by large, powerful entities (such as the government of Togo) (N = 11 of 15). 

\subsubsection{Aspects of Mobile Phone Metadata Considered Sensitive}
Participants primarily felt that mobile phone metadata was sensitive to the extent that it could reveal something sensitive about their finances. In particular, data about mobile money account balances and transactions were the pieces of mobile phone metadata that people felt were most sensitive because this type of data directly revealed finances. Other data contained in mobile phone metadata --- such as data about phone calls, text messages, and location --- was less commonly raised as sensitive, unless participants perceived that it could be used to infer something sensitive about their finances. For example, several participants mentioned that placing a lot of phone calls or traveling throughout the country could indicate that a person’s business is going well. Likewise, not placing a lot of phone calls or being the one to receive mobile money transfers could reveal that finances were not going well. 

While much of the interview focused on the potential for mobile phone metadata to reveal financial data, we did explicitly ask participants about the sensitivity of other pieces of data that are contained in mobile phone metadata: specifically location \cite{de2013unique, kohli2023privacy} and social contact networks \cite{eagle2008mobile, eagle2009inferring} (motivated by the fact that much of the research in the literature by domain experts has focused on the privacy harms from making data on location traces and contact networks available \cite{2012us, hosein2013aiding, mayer2016evaluating, thompson2019twelve, ovide2021}). However, our participants generally did not find this sort of data sensitive unless they could envision how it could be misused or used to infer information that otherwise felt sensitive. For example, Soumbalo (a 38 year old man), mentioned that he would not want his wife to see his mobile money transactions because it would reveal that he was cheating on his wife. However, records related to phone calls or text messages --- including who he was speaking to --- would not raise the same set of concerns. When asked why mobile money transactions would reveal infidelity but phone call records would not, he said, ``when it's money, she knows that you are exchanging something with that person. With money, it's clear that there's something. But for the phone call, [my wife] doesn't know the content of the conversation.'' Soumbalo felt that data about mobile money transactions would be interpreted in a way that data about phone calls could not. 

Although we did not directly ask participants how they could envision mobile phone metadata as beneficial, some (N = 4 of 15) participants’ initial reaction was that they were ``happy'' to learn that this data was recorded because the records could help settle interpersonal disputes. For example, if one person said they transferred mobile money and another person said they did not receive it, mobile phone metadata could provide an accurate record. Moreover, while we did not ask participants to specifically reflect on the sensitivity of communication content versus communication metadata, several participants drew a distinction: the content of phone calls or text messages felt much more sensitive than the metadata because the content contained ``secrets.'' 

\subsubsection{Mobile Phone Metadata and Shame}
Participants occasionally perceived that mobile phone metadata could reveal moments of financial hardship: causing shame if shared with people in their household or people in their village. It was often instances where data containing mobile money account balances or transactions records could reveal financial hardship. For example, Soumbalo said that it would be a problem if mobile money balances and transactions were made available to people in his village because it would reveal who was doing well (leading to jealousy) and who was not doing well (leading to shame). He said, ``the one who gives [mobile money] can be harmed if it is known in the community because it is known that he has money and gives it to others. Some may be jealous of him. But the one receiving may feel some kind of shame: people in the community will have a bad opinion of him because he depends on other people to survive.'' Here we see how people would be concerned if mobile phone metadata was shared with people in their household or village  --- particularly aspects that indicated financial hardship --- because it could bring shame.

\subsubsection{Mobile Phone Metadata and Jealousy}  
More common than exposing moments of financial hardship, participants often perceived that mobile phone metadata could reveal moments of financial success: causing jealousy if shared with people in their household or people in their village. For example, Soumbalo said he would not want people in his village to know the amount of money on his mobile money account because, ``what one possesses in their pocket should not be known to others. If a wicked person knows the amount of money you have, they can kill you...you don’t know who is for or against your success in life. That is the reason we should keep our goods secret.''  Beyond mobile money transaction records, Senyo shared how other people might interpret data about his travel to suggest that his business is doing well, which could similarly result in envy leading people to ``do something that can affect my life.'' Similarly drawing on the ways that different aspects of mobile phone metadata could reveal financial success, Tchilalo said that it would be a problem if people in her village knew the data recorded about phone calls or location because both could be used to infer if her business was doing well. If people knew her business was going well, they may become jealous and have ``bad intentions on you, the person [may] harm you spiritually and kill you.'' As the above examples illustrate, mobile phone metadata --- both metadata specifically about mobile money, as well other transaction records --- could be used to draw inferences about financial success, which may lead  others nearby to become jealous and invoke witchcraft. 

\subsubsection{Mobile Phone Metadata and Autonomy}  
Participants perceived that the mobile phone metadata could reveal financial data that, if shared with spouses, could limit their autonomy and control over money. For example, when asked how he would feel if his wife knew about his mobile money transactions, Soumbalo said, ``Here, there exists a two-sided reasoning. There are some people whose wives know about the money their husbands possess; and there are some men whose wives do not. It depends on the mood in the household. Suppose a man transfers money to relatives of his wife or someone in his own family to solve a specific problem. [If there is a healthy relationship] it is a good thing because the wife is aware of the necessity. On the contrary, if there exists an unhealthy relationship between the wife and husband, then the wife can protest [because she believes the money was not spent responsibly].'' Beyond accusations of using money foolishly, Awete (a 46 year old man) said he was concerned that sharing mobile money balances with his wife could be a problem because ``she would steal the money and run away,'' drawing on a recent experience where ``one of my friends went to work and when he came back home, his wife left with everything they had at home except the house. He doesn’t know where she has gone. You see, women are dangerous if things are not going well at home.'' Padayodi (a 30 year old woman) similarly expressed a concern that her husband could steal the money saying, ``if I get along very well with my husband then he can know whatever I have [on my mobile money account], but if we do not have a good relationship I wouldn't let him know about my finances because if he knows I have money he will take the money and me and my kids will suffer.'' 

\subsubsection{Mobile Phone Metadata and The Privacy Protection of Distance} While participants expressed concerns about the pieces of data recorded in mobile phone metadata that could reveal sensitive financial information, participants were largely not concerned if the identical data was shared with entities that were socially and physically distant, such as the government of Togo. For example, Soumbalo mentioned earlier that he would not want people in his village to know about mobile money transactions because of the shame and jealousy it would bring to both the person giving money and the person receiving money; he also said that he would not want his wife to know about mobile money transactions because she could find out about infidelity. When asked how he would feel about the government of Togo knowing this same data, Soumbalo said, ``I won’t be disturbed if the Togo government knows about [mobile money balances and transactions]. They know that you are helping others. But among ourselves here, it may be disturbing.'' When asked to reflect on why it was fine for the government to know this data but not people nearby, he said, ``the reason why the government could know about it is that they can’t envy me.'' Here we see that it was the people nearby --- the people in his household or village --- that he was concerned might have access to this data because they might use it against him in a way that would not be possible if it was accessed by the government, an entity that he perceived to be both physically and socially distant. Tchilalo, who earlier mentioned that she would not want people in her village to know data recorded about phone calls and location because they may become jealous of her business's success, said that she ``would not be disturbed'' if the government of Togo had access to this data because ``first, they are far from me and second they don’t know me in person... the government cannot harm you or [become] jealous, but people living in the village can harm you or be jealous of you.'' These quotes suggest that the physical and social distance of the government implied that it could not make someone feel shame, become jealous, or limit an individual’s autonomy. 

Interestingly, prior research in Namibia \cite{VanStaden2023LocalizsedTrust} and India \cite{Srinivasan2018povertyprivacy} on digital IDs found that people tended to develop relationships with organizations (including the state) that evolved over time and shaped their perception of privacy. While we did not ask participants to talk about their prior relationships with the government, participants did not often talk about privacy concerns in terms of their relationship with the state, whereas relationships were central in their discussion of privacy harms that could arise from people nearby. 

While it was not common in our data, one participant raised concerns that mirrored some of the concerns of domain experts, particularly related to the potential for new data and technologies to enable a surveillance state. After explaining data recorded about mobile money account balances and transactions, we asked Djidula (a 60 year old man) how sensitive (or not) this data felt. He responded that the data felt sensitive because, ``if you have a phone, you are under their control. It is like your father. He gives birth to you and tells you this is what to do, for instance, you have to use your phone to use Flooz or T-Money [the two mobile money providers in Togo], so you are under your father, you are under his control, you have no other way out.'' He felt that this meant that, ``[the mobile phone] is watching what you are doing...so he who has a phone has nowhere to hide.'' After explaining data recorded about phone calls, Djidula said, ``before the modern era, things can be hidden but with a phone, nothing can be hidden'' and went on to provide the example of money: ``in the old times, men for instance, they can hide money somewhere and nobody will know, they will only discover the money after his death. Things have changed and it is now related to phones. If you have a phone, whatever you do here in this village, they know.'' In this interview, Djidula expressed a set of concerns that envisioned large, powerful actors that are far away surveilling his behavior. This concern --- that once the data is recorded, there is a sense of being ``watched'' with ``nowhere to hide'' --- is one of the common concerns raised by domain experts around the use of big data approaches to development policy \cite{taylor2015name, taylor2016no}. 

\subsubsection{Mobile Phone Metadata: Responses to ``Harmful'' and ``Helpful'' Scenarios}
In the third part of the interview, we asked participants (N = 13) to reflect on four scenarios about how mobile phone metadata could be used in practice by the government of Togo (see Appendix \ref{app:scenarios} for the scenarios). In brief: In \textit{Scenario 1 --- helpful wealth} the government of Togo wants to give money to the poorest people in the country. The government decides to use mobile phone metadata to determine who is wealthy or poor. In \textit{Scenario 2 --- harmful wealth} the government of Togo wants to improve the economy and has decided to force the poorest people to work. The government decides to use mobile phone metadata to determine who is wealthy or poor. In \textit{Scenario 3 --- helpful religion} the government of Togo wants to sensitize information about a new government program at a time of day when people will be available, not attending religious services. The government decides to use mobile phone metadata to determine if someone is actively practicing religion. In \textit{Scenario 4 --- harmful religion} the government of Togo wants to punish people who are not actively practicing a religion. The government decides to use mobile phone metadata to determine if someone is actively practicing religion. As discussed in Section \ref{sec:methods}, the four scenarios were inspired by examples domain experts use to argue for, or against, the use of big data in development. Scenarios were read aloud to participants and the order of the scenarios was randomized. 

We find that all of the participants we interviewed did not want mobile phone metadata to be used for purposes that they perceived to be harmful (i.e., punishing people for not practicing religion or forcing people to work); and that they did want mobile phone metadata to be used for uses that they perceived to be helpful (i.e., providing cash assistance to the poorest people or timing important messages to sensitize government programs). In the case of ``harmful'' scenarios, participants' concerns were often not about the use of mobile phone metadata per se, but instead with the particular use case. For example, after reading \textit{Scenario 2 -- harmful wealth} (described above and in Section \ref{app:scenarios}), we asked Naka (a 30 year old woman) if she had any initial reactions to the scenario. She responded, ``this question is difficult for me to answer because it is a kind of slavery that’s coming back in Togo.'' In the case of ``helpful'' uses, while participants typically wanted mobile phone metadata to be used to support the scenarios presented, many (N = 10 of 13) situated the mobile phone metadata in their lives to point out instances whereon the ground uses of mobile phones could complicate its ability to generate accurate inferences. For example, after reading \textit{Scenario 1 --- helpful wealth} (described above and in Section \ref{app:scenarios}), Soka (a 40 year old woman), expressed that data recorded in mobile phone metadata may not always be an accurate representation of who has money because, ``a person can have money on their mobile money account but the money does not belong to them so...[mobile money account balances] don't tell reality.'' We would often followup by asking participants if they had any other concerns beyond accuracy and/or if accuracy was not an issue, would they have any other concerns. In all cases, participants confirmed that accuracy was their only concern. To summarize: our participants did not want mobile phone metadata to be misused in ways that they perceived to be harmful; but, participants wanted mobile phone metadata to be used in ways that they perceived to be helpful, so long as the mobile phone metadata and the inferences were accurate. Notably, concerns about surveillance, power, data sharing, and control were largely absent.

\section{Discussion}
We begin by discussing one point of convergence between experiential expert and domain experts. Both tend to agree that big data should be used to enable ``helpful'' uses while preventing ``harmful'' ones. The use of a scenariobased approach enabled us to identify that participants wanted the government to use mobile phone metadata in ways that are ``helpful'' (i.e., providing cash assistance to the poorest people or timing important messages to sensitize government programs) but have concerns about ``harmful'' uses (i.e., punishing people for not practicing religion or forcing people to work). Privacy and development scholars tend to agree that ``helpful'' uses should be supported and ``harmful'' ones mitigated \cite{landau2016transactional, mann2018left, Abebe_2021}, though there is some disagreement within the scholarly community about how to appropriately balance ``helpful'' and ``harmful'' uses \cite{sandvik2017no, de20198}.

While our experiential and domain experts tend to agree that ``helpful'' uses should be supported ``harmful'' ones should be avoided, they diverge on the salient ``harmful'' uses. The experiential experts we interviewed tended to take a relational view; situating mobile phone metadata in everyday life, our experiential experts raise privacy concerns related to data being shared with people nearby (i.e., spouses, household, and neighbors). In contrast, domain experts tend to focus on privacy concerns related to surveillance and misuse by powerful actors who are far away \cite{de2013unique, taylor2016no, landau2016transactional, Abebe_2021}.

Moreover, the relational privacy harms surfaced by our experiential experts tended to center relational harms related to shame, jealousy, and autonomy. In contrast, domain experts typically surface less context-specific privacy harms, such as surveillance \cite{taylor2016no}, (mis)using data to target individuals or groups \cite{landau2016transactional, taylor2016no}, and extracting profit in ways that mirror patterns of colonization \cite{Abebe_2021}. Taken together, the privacy harms that our experiential experts surfaced are largely ``out of view'' to domain experts, and vice versa.  

The different privacy concerns raised by the two types of experts can likely be explained in part by the different types of expertise each brings to bear, as well as local norms and expectations of privacy that differ across cultures and contexts. Experiential experts are particularly well-positioned to surface privacy harms that arise when a new technology is introduced into their everyday lives: carefully managing what data is shared about them, and with whom (among others). Domain experts, on the other hand, are particularly well-suited to identify technical vulnerabilities, situate a given technology in history, and identify interactions between technology and policy (among others).\footnote{Our results parallel previous work on the introduction of new location tracking applications in the US. In that setting, privacy and security experts surfaced privacy concerns related to corporate and government surveillance, while experiential experts surfaced privacy concerns related to intimate partner violence \cite{Slupska2019IPV, Levy2020IPV}.} Beyond expertise, our results may also be explained by the important ways that privacy norms differ across cultures and contexts \cite{Abokhodair2016Privacy, Srinivasan2018povertyprivacy, Jack2019Cambodia}. Most scholarship on data privacy has come from research conducted in ``western'' countries --- which tends to view privacy as individualizing --- whereas our experiential experts are in a ``non-western’’ country which tends to view privacy as more collectivist and relational \cite{Vashistha2018HCI4D, VanStaden2023LocalizsedTrust}. 

The findings from this research should not be interpreted to suggest that the divergence between these experiential experts and domain experts is a problem that needs to be rectified. Instead, the privacy concerns surfaced by each group of experts are more robust when placed together: each surfacing a set of privacy concerns that are ``out of view'' to the other. Bringing these perspectives together enables a more holistic account of privacy, positioning technologists and policymakers to more constructively address such privacy concerns. In the final section of this paper we take a first step toward providing design implications for the use of big data approaches to development policy that integrate the privacy concerns raised by both experiential experts and domain experts.

\section{Design Implications}
\label{sec:design}
We have documented how the privacy concerns raised by experiential experts tend to diverge from those commonly raised by domain experts. In what follows, we use two case studies to demonstrate how bringing together the privacy concerns raised by both sets of experts can be more robust and constructive than either one on its own. This section is a first step in a larger effort to take seriously the privacy concerns surfaced by different types of experts, and highlights the need for future work to consider how to elicit, assemble, and design for more holistic accounts of privacy.

\subsection{Data Sovereignty Models}
The data governance community has begun to explore alternative models of data sovereignty including data trusts, data unions, and data cooperatives to address concerns related to consent, surveillance, and the centralization of power that arise from traditional data ownership models \cite{duncan2023data, olorunju2023african}. While these models differ in their specific implementations, they are largely oriented toward shifting ownership and control of data to people and communities. If designers only paid attention to the privacy concerns raised by domain experts, this would likely result in designs that minimize harms related to consent, surveillance, and unequal power relations. For example, one reasonable design might provide data directly to data subjects or a representative of the community (such as a village elder). However, such a design would not address the issues of central concern to our experiential experts: shame, jealousy, and limited autonomy that arise when data is shared with people nearby. On the other hand, if designers only paid attention to the privacy concerns raised by experiential experts, this would likely result in designs that minimize harms related to relational privacy: shame, jealousy, and limited autonomy. For example, one reasonable design could provide data only to governments or corporations (perhaps not unlike common data ownership models that exist today). While such a design might better address relational privacy concerns, it would likely raise issues related to consent, surveillance, and unequal power relations. Any design that only considers the privacy harms raised by one set of experts is thus incomplete. 

In contrast, placing the two in conversation outlines a more robust and constructive design space: data sovereignty models that both provide people and communities with more control over their data \textit{and} safeguard against relational privacy harms. For instance, one could imagine a data sovereignty model that pairs local data storage alongside a strong data governance framework that limits who in the community can access data and for what proposes, as well as comprehensive sunset clauses on the collection, storage, analysis, and sharing of data to limit future misuse by large, powerful actors.

\subsection{Poverty Prediction for Aid Allocation}
As a second example, we consider the original context that led us to Togo: the use of mobile phones and mobile phone metadata to determine eligibility for an anti-poverty program \cite{aiken2022machine}. If program designers only paid attention to the privacy concerns commonly raised by domain experts (see Section \ref{sec:domainexperts}), they might implement data-centric privacy approaches (such as differential privacy \cite{dwork2019differential} or k-anonymity \cite{sweeney2002k}) to de-identify the underlying data, leverage data sovereignty models to address concerns of surveillance, or develop methods to obtain consent. While such designs might address the privacy harms raised by domain experts, they would likely not address the types of relational privacy harms of central concern to our experiential experts. For example, the information displayed on mobile interfaces in the delivery of aid could inadvertently reveal sensitive information about poverty and beneficiary status to household and community members, particularly in the context of ubiquitous device sharing in low-income contexts \cite{Burrell2010PhoneSharing, Ahmed2017PhoneSharing, Aiken2022PhoneSharing}. On the other hand, if the program designers only paid attention to these types of relational privacy concerns raised by experiential experts, solutions may focus on designing interfaces that enable beneficiaries to receive cash transfers on shared mobile phones in ways that do not reveal one's beneficiary status \cite{Ahmed2019PersonalStuff}, develop digital literacy programs to enable people to use devices on their own, or focus on program messaging to limit the shame and stigma associated with receiving aid. However, only paying attention to these relational privacy concerns might unintentionally enable surveillance or data misuse by government actors or private corporations. Again, a design that only considers the privacy harms raised by one set of experts is incomplete. 

Here too, placing the two in conversation outlines a more robust and constructive design space: using mobile phones and digital data to allocate cash aid in ways that safeguard against future misuses \textit{and} minimize relational privacy harms. For instance, one could imagine pairing comprehensive sunset clauses to place limits on who can access underlying mobile phone metadata, for which purposes, and for how long alongside interface designs that do not ``out'' beneficiaries to reduce stigma and/or use visuals in interface design to enable people with lower literacies to be able to register and access cash aid on their own.  

More generally, these two case studies illustrate the importance of taking seriously the privacy concerns raised by both domain experts and experiential experts: each bringing a set of privacy concerns ``into view'' that are largely absent from the other.

\section{Conclusion}
In this paper, we sought to understand the data privacy concerns raised by experiential experts about the use of big data (in particular, mobile phone metadata) in development policy. We conducted semi-structured interviews with people living in rural villages in Togo, finding that the primary data privacy concerns of experiential experts were different than those of domain experts in privacy and development. Rather than focusing on surveillance and the centralization of power, the primary concerns of experiential experts were relational, often tied to notions of jealousy, shame, and autonomy. We demonstrate how placing the privacy harms raised by experiential experts and domain experts in conversation can provide a more robust and constructive account of privacy than either on its own. We hope our efforts can help inspire future work that explores how to better elicit the privacy concerns and values of a wider range of ``experts,'' and to better position technologists and policymakers to design technologies, policies, and practices that integrate insights from different types of expertise to address holistic conceptions of data privacy.



%
\newpage
\bibliographystyle{ACM-Reference-Format}
\bibliography{references}

\newpage
\appendix

\section*{Appendices}

\section{Participant Table}
\label{app:participant_table}

\begin{table}[H]
\caption{Summary statistics on interview participants, together and broken down by interview phase.}
\scriptsize
\begin{tabular}{lcccccc}
\toprule
                               & \textbf{Phase 1} & \textbf{Phase 2} & \textbf{Phase 3 } & \textbf{Phase 3 } & \textbf{Phase 3 } & \textbf{Total} \\ 
                               &  &  & \textbf{(village 1)} & \textbf{(village 2)} & \textbf{(village 3)} & \\ 
                               \midrule
\multicolumn{7}{l}{\textit{Panel A: Gender}}                                                                                                                                       \\
Male                           & 7                & 9                & 17                           & 25                           & 15                           & 73             \\
Female                         & 9                & 11               & 10                           & 11                           & 10                           & 51             \\ \\
\multicolumn{7}{l}{\textit{Panel B: Age}}                                                                                                                                          \\
35 years older or less         & 7                & 11               & 9                            & 10                           & 14                           & 51             \\
Older than 35 years            & 8                & 9                & 17                           & 19                           & 10                           & 63             \\
Not reported                   & 1                & 0                & 1                            & 7                            & 1                            & 10             \\ \\
\multicolumn{7}{l}{\textit{Panel C: Occupation}}                                                                                                                                   \\
Farmer                         & 8                & 13               & 23                           & 20                           & 16                           & 80             \\
Other                          & 8                & 7                & 4                            & 16                           & 9                            & 44             \\ \\
\multicolumn{7}{l}{\textit{Panel D: Education}}                                                                                                                           \\
None                           & 1                & 6                & 7                           & 14                           & 8                            & 36             \\
Completed part or all of primary school                            & 5                & 3                & 4                            & 5                            & 7                            & 24              \\
Completed part of all of middle school                           & 3                & 4                & 11                           & 14                           & 8                            & 40             \\
Completed part or all of high school                     & 7                & 6                & 5                            & 1                            & 2                             & 21             \\
Completed all or part of university                        & 0                & 0                & 0                            & 2                            & 0                            & 2              \\
Did not ask                        & 0                & 1                & 0                            & 0                            & 0                            & 1              \\\\
\multicolumn{7}{l}{\textit{Panel E: Ethnic group}}                                                                                                                                 \\
Akebou                         & 1                & 0                & 0                            & 0                            & 0                            & 1              \\
Akposso                        & 1                & 0                & 0                            & 0                            & 0                            & 1              \\
Ewe                            & 2                & 11               & 0                            & 0                            & 24                           & 37             \\
Fon                            & 1                & 0                & 0                            & 0                            & 0                            & 1              \\
Gangan                         & 1                & 0                & 0                            & 0                            & 0                            & 1              \\
Kabyé                          & 2                & 9                & 27                           & 18                           & 1                            & 57             \\
Kotokoli                       & 1                & 0                & 0                            & 10                           & 0                            & 11             \\
Laffa                          & 1                & 0                & 0                            & 0                            & 0                            & 1              \\
Lamba                          & 2                & 0                & 0                            & 0                            & 0                            & 2              \\
Moba                           & 3                & 0                & 0                            & 0                            & 0                            & 3              \\
Peule                          & 1                & 0                & 0                            & 8                            & 0                            & 9              \\ \\
\multicolumn{7}{l}{\textit{Panel F: Religion}}                                                                                                                                     \\
Animist                        & 1                & 3                & 11                           & 6                            & 11                            & 32             \\
Christian                      & 11               & 15               & 15                           & 8                            & 14                           & 63             \\
Muslim                         & 3                & 0                & 0                            & 19                           & 0                            & 22             \\
None                           & 1                & 2                & 1                            & 3                            & 0                            & 7              \\ \\
\multicolumn{7}{l}{\textit{Panel G: Interview language}}                                                                                                                           \\
Ewe                            & 2                & 11               & 0                            & 0                            & 25                           & 38             \\
French                         & 10               & 0                & 0                            & 0                            & 0                            & 10             \\
Kabyé                          & 1                & 9                & 27                           & 19                           & 0                            & 56             \\
Kotokoli                       & 3                & 0                & 0                            & 17                           & 0                            & 20             \\ \\
\multicolumn{7}{l}{\textit{Panel H: Novissi beneficiary status}}                                                                                                                   \\
Beneficiary                    & 6                & 1                & 13                           & 26                           & 5                            & 51             \\
Registered but did not receive & 5                & 1                & 13                           & 8                            & 15                           & 42             \\
Unregistered                   & 5                & 0                & 1                            & 2                            & 5                            & 13             \\
Did not ask                    & 0                & 18               & 0                            & 0                            & 0                            & 18             \\ \bottomrule
\end{tabular}
\end{table}

\newpage

\section{Mobile Phone Metadata in Development Policy}
\label{app:mobilephonemetadata}
Mobile phone metadata are held by mobile network operators, and have on occasion been shared with governments, nonprofit organizations, for-profit organizations, and researchers to inform development policy decisions \cite{taylor2015name, milusheva2021challenges}. For example, mobile phone metadata have been used to map poverty in Rwanda \cite{blumenstock2015predicting}, Afghanistan \cite{blumenstock2018estimating}, Guatemala \cite{hernandez2017estimating}, and Bangladesh \cite{steele2017mapping}; inform the targeting of humanitarian aid in Togo \cite{aiken2022machine} and the Democratic Republic of the Congo \cite{mukherjee2023digital}; measure mobility in response to natural disasters in Haiti \cite{bengtsson2011improved} and Nepal \cite{wilson2016rapid}; and predict the spread of disease in Kenya \cite{wesolowski2013use},  Sierra Leone \cite{peak2018population}, and Senegal \cite{milusheva2020managing}. In Togo, mobile phone metadata was as part of the \textit{Novissi} program that targeted cash assistance to the poorest people in the country to help them survive economic impacts caused by pandemic shutdowns \cite{aiken2022machine}. 

While the exact information recorded in mobile phone metadata varies between operators, the mobile phone metadata analyzed in the context of Togo’s \textit{Novissi} program are broadly emblematic of the types of information typically recorded by mobile network operators about transactions placed on mobile networks. These include: 

\textbf{For calls and text messages. } Phone number of the caller and recipient, time of the transaction, duration (for calls), and location of the cell tower through which a call is placed. 

\textbf{For mobile data usage.} Time and amount of mobile data usage.\footnote{Unlike in the data recorded by smartphone apps, mobile phone metadata recorded by mobile network operators does not include information about the exact locations of phone users, only information about the locations of cell towers through which calls are placed and received. This information is only recorded when the user uses their phone (phone call, text message, mobile money transaction, etc.).} 

\textbf{For mobile money usage.} Mobile money is an approach to storing and transferring money available on both basic and smartphones in many LMICs, including Togo. Both of Togo's two mobile network operators provide mobile money platforms. For mobile money transactions, the following information is recorded: phone number of the person transferring and receiving money, time of the transaction, type of transaction (including cash-in, cash-out, peer-to-peer transfers, and purchases from registered merchants), amount of money transferred, and total balance of the sender and recipient.

\section{Additional methodological details} 

\subsection{Description of Villages - Fieldwork}
\label{app:villagedescription}
Fieldwork was conducted in three rural villages in Togo. We provide a brief description of each village below. 

\textbf{Village 1.} In the northern part of the country, Village 1 is about a 40 minute drive on a paved road from Kara, the second largest city in Togo. People are largely Kabyé and speak Kabyè, practicing Christianity, Animism, and to a smaller degree Islam. The Canton was eligible for \textit{Novissi} and targeting was at an individual level based on machine learning and mobile phone data as described in Aiken et al. 2022 \cite{aiken2022machine}. The percent of eligible voters who registered for \textit{Novissi} was comparatively quite high (37.32 percent). 

\textbf{Village 2.} In the middle part of the country, Village 3 is about a 1 hour drive on a mix of dirt and paved roads from Kara. People are a mix of Kabyé, Kotokoli, and Peule/Fulani, practicing Christianity, Animism, and Islam. Unlike the two other villages, everyone in Village 2 -- and the entire canton -- was eligible to receive benefits from \textit{Novissi}. Everyone living in the canton of Soudou, which Village 2 is in, was eligible for a one-time \textit{Novissi} cash transfer in July 2020, due to movement restrictions associated with an outbreak of COVID-19 in the region. We specifically recruited this canton because it was the one canton in Togo where there was regional-targeting: where everyone registered to vote in the Canton was eligible to receive benefits. While sampling was done with this in mind, differences based on targeting approach do not show up in this paper because perceptions of data privacy did not differ based on the targeting approach. 

\textbf{Village 3.} In the southern part of the country, Village 3 is about a 1 hour drive on a dirt road from Tsevié, a suburb of the capital city of Lomé. People are largely Ewe, practicing Christianity or Animism. The Canton was eligible for Novissi and targeting was at an individual level based on machine learning and mobile phone data as described in Aiken et al. 2022 \cite{aiken2022machine}. The percent of eligible voters who registered for Novissi was comparatively quite low (0.83 percent).

\subsection{Scenarios Read Aloud to Participants in Part 3 of Interviews}
\label{app:scenarios}

Note: In interviews, the order the scenarios were read was randomized.

\medskip

\noindent \textbf{Scenario 1 -- Helpful wealth:} I want you to imagine for a moment that the government of Togo is concerned about people in the country who are poor. To improve the lives of people, the government has decided to give money to the poorest people. To identify who will receive money and who will not, the government needs to find a way to identify who in Togo is the poorest. They decide to use information recorded by mobile phones. A person might be wealthier if they have a lot of mobile money in their mobile money account, or because they travel a lot to Lomé. Using the information recorded by mobile phones, a person would likely receive money if they were poor, and not if they were wealthy. 

\medskip

\noindent \textbf{Scenario 2 -- Harmful wealth:} I want you to imagine for a moment that the government of Togo is concerned about the economy. To improve the economy, the government has decided that it will force the poorest people to work in the country. To identify who will be forced to work and who will not, the government needs to find a way to identify who in Togo is the poorest. They decide to use information recorded by mobile phones. A person might be wealthier if they have a lot of mobile money in their mobile money account, or because they travel a lot to Lomé. Using the information recorded by mobile phones, the government would likely force a person to work if they were poor, and not if they are wealthy.

\medskip

\noindent \textbf{Scenario 3 -- Helpful religion:} I want you to imagine for a moment that the government of Togo wants to improve the way it sensitizes information about new government programs. To improve its sensitization of programs, the government has decided to call peoples’ mobile phones. The government wants to make sure that phone calls do not arrive at a time when someone is attending a religious service. To avoid calling people during a religious service, the government needs to find a way to identify who in Togo is actively practicing religion and who is not. They decide to use information recorded by mobile phones. A person might be religious if they do not place calls or answer calls during muslim prayer times or church services. Using the information recorded by mobile phones, the government would likely place phone calls to people when they are not attending religious services, and not place phone calls when they are attending religious services. 

\medskip

\noindent \textbf{Scenario 4 -- Harmful religion:} I want you to imagine for a moment that the government of Togo is concerned that people are becoming less religious. To increase religiosity in the country, the government has decided to punish people who do not actively practice religion. To identify who will be punished for not practicing religion and who will not, the government needs to find a way to identify who in Togo who are practicing religion. They decide to use information recorded by mobile phones. A person might be religious if they do not place calls or answer calls during muslim prayer times or church services. Using the information recorded by mobile phones, the government would likely punish a person if they were not practicing the religion devoutly, and not punish a person if they were practicing devoutly. 

\newpage


\end{document}